\documentclass{Rinton-P9x6}

\begin{document}

\title{To appear in the Proceedings of the 6th International
Conference on \\{\it Quauntum Communication, Measurement and
Computing} \\(QCMC 2002), Boston, Massachusetts, July 22-26,
2002. \\
\vspace{1 in}
Taming Entanglement}

\author{Paul Kwiat, Joe Altepeter, David Branning, Evan Jeffrey,
Nicholas Peters, and Tzu-Chieh Wei}


\maketitle

\abstracts{ Using a spontaneous parametric-downconversion source of
photon pairs, we are working towards the creation of arbitrary 2-qubit
quantum states with high fidelity.  Currently, all physically
allowable combinations of polarization entanglement and mixture can be
produced, including maximally-entangled mixed states.  The states are
experimentally measured and refined via computer-automated
quantum-state tomography, and this system has also been used to
perform single-qubit and ancilla-assisted quantum process tomography.
}


\begin{figure}[b!]
\epsfxsize=24pc 
\begin{center}
\epsfbox{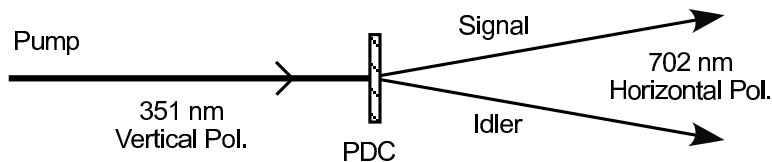} 
\end{center}
\caption{Spontaneous parametric downconversion.}
\label{SinglePDC}
\end{figure}

Central to the long-term future of quantum information processing is
the capability of performing extremely accurate and reproducible gate
operations.  The restrictions for fault-tolerant quantum computation
are extremely demanding: the tolerable error-per-gate operation should
be less than $10^{-4}$ to $10^{-6}$.  Implementing such precise gate operations
and preparing the requisite input states is therefore one of the key
milestones for quantum information processing.  Using optical
realizations of qubits, e.g., polarization states of photons, we have
the potential to meet these demanding tolerances.  Therefore, although
large-scale quantum computers will perhaps never be constructed solely
using optical qubits, these systems nevertheless form a unique and
convenient testbed with which to experimentally investigate the issues
surrounding state creation, manipulation, and characterization, and
also ways of dealing with decoherence.

Our primary tool for these investigations is a source of correlated
photons produced via the process of spontaneous parametric down
conversion: with small probability, a pump photon of appropriate
polarization may split into two longer-wavelength daughter photons,
subject to energy and momentum conservation (Fig.~\ref{SinglePDC}).
By triggering on one of these photons, the other is prepared in a
single-photon Fock state\cite{Mandel}.  We can apply local unitary
transformations to the polarizations of these photons using a
birefringent half-waveplate (HWP) and quarter-waveplate (QWP).  We can
also introduce decoherence (either independently or collectively) by
passing the photons through birefringent delay lines.

\begin{figure}[t]
\epsfxsize=28pc 
\begin{center}
\epsfbox{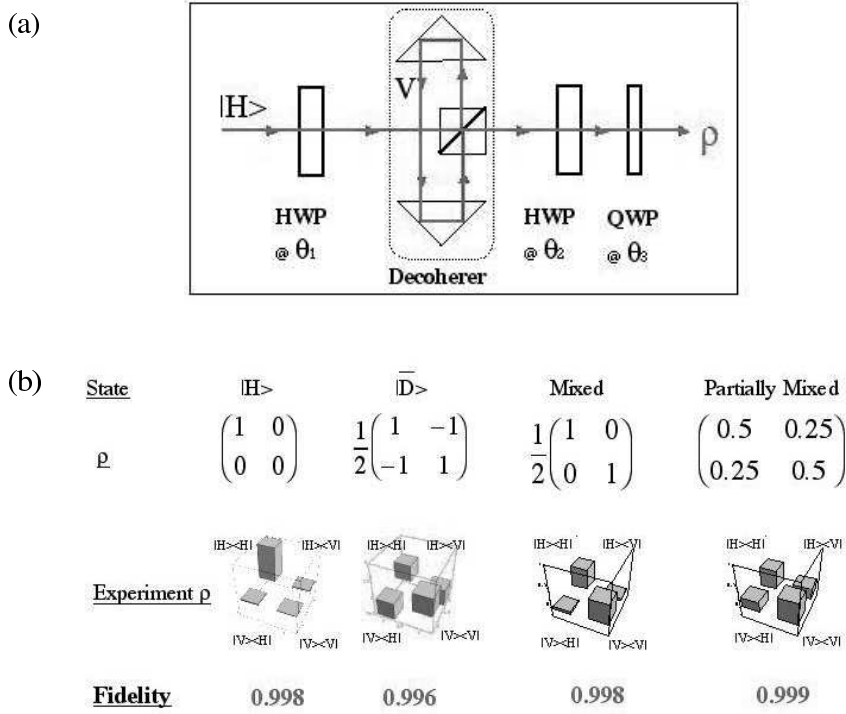} 
\end{center}
\caption{(a) Experimental arrangement for single-photon state
creation.  The single-photon input to the system comes from one member
of a parametric downconversion pair, with the other photon used as a
herald. The operation of the decoherer -- to separate orthogonal
polarizations by much more than the coherence length -- is indicated
here using a polarizing beamsplitter to create a birefringent delay
line. (b) A variety of single-qubit states have been generated and
reconstructed using quantum state tomography.}
\label{SingleQubit}
\end{figure}

\begin{figure}[t]
\epsfxsize=26pc 
\begin{center}
\epsfbox{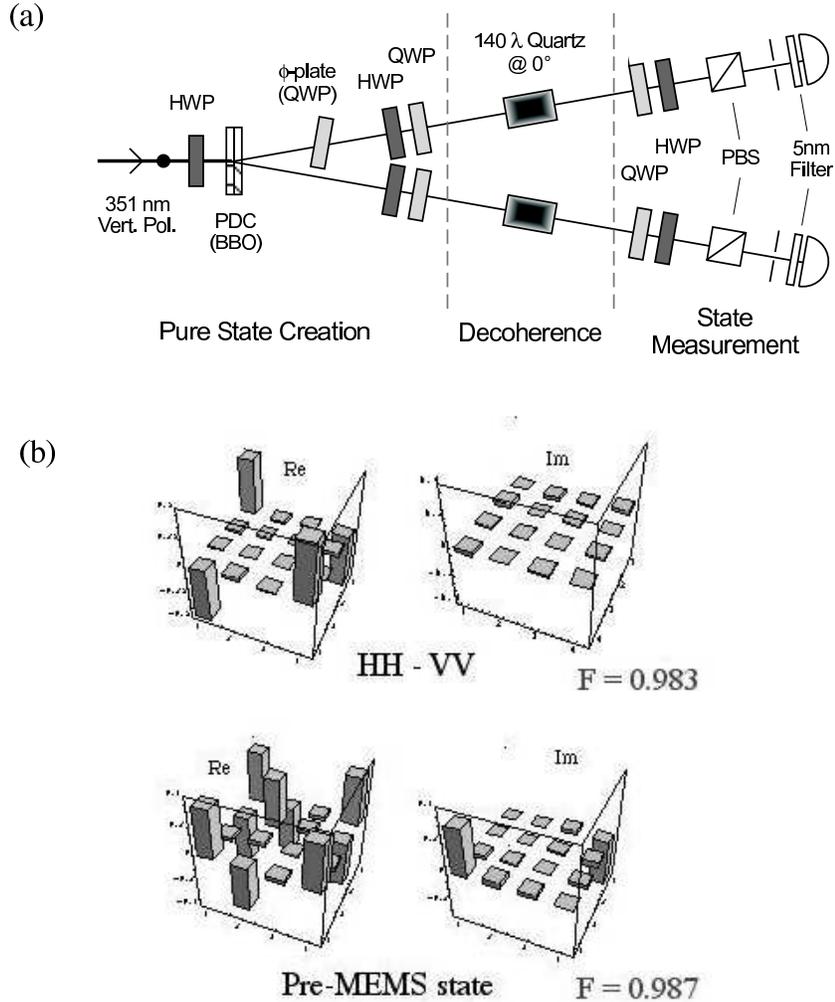} 
\end{center}
\caption{(a) Experimental arrangement for two-photon state creation.
(b) A variety of two-qubit states have been generated and
reconstructed using quantum state tomography. Also shown are the
fidelity of the reconstructed density matrices with the target input
states, indicating a high degree of control.}
\label{TwoQubit}
\end{figure}

\begin{figure}[b]
\epsfxsize=26pc 
\begin{center}
\epsfbox{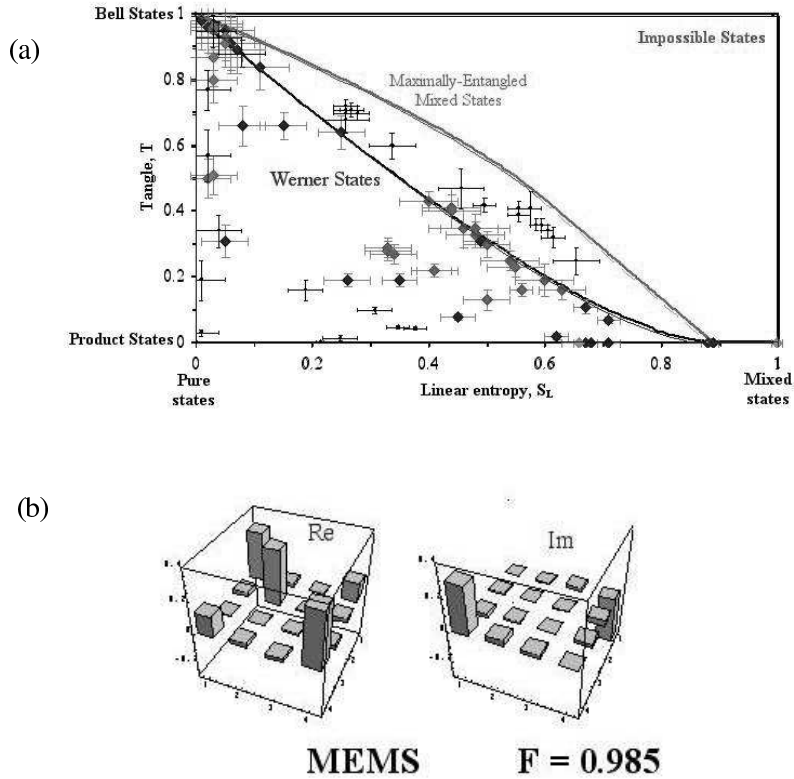} 
\end{center}
\caption{(a) The Tangle-Entropy plane. (b) Tomographic reconstruction
of the density matrix for a MEMS (Maximally Entangled Mixed State).
F is the fidelity of the measured state with the target.}
\label{TangleEnt}
\end{figure}

Using these techniques for the single photon case, the initial pure
horizontal state $|H\rangle$ may be precisely converted into an
arbitrary pure or mixed state (Fig.~\ref{SingleQubit}).  We estimate
that we can create and reliably distinguish (with fidelities of 0.998
or better) over 100,000 single-qubit states.  Applying these
single-qubit techniques to each output of a downconversion crystal, we
can create arbitrary {\it product} states for the two photons.  But
these comprise only a very small part of the total two-qubit Hilbert
space.  To access the rest, we must create entangled states; this is
done by adding a second downconverter with an orthogonal optic axis as
shown in Fig.~\ref{TwoQubit}a.  A given pair of signal and idler
photons could have been born in the first crystal, with vertical
polarizations, or in the second with horizontal polarizations.  These
two possibilities cannot be distinguished by any measurements other
than polarization, so the quantum state for these photons is a
superposition of $|V\rangle|V\rangle$ and $|H\rangle|H\rangle$.
Because each crystal responds to only one pump polarization, the
relative weights of the two downconversion processes can be controlled
by adjusting the input pump polarization\cite{KwiatSource}.  A
birefringent phase plate is also added to one of the outputs to
control the relative phase of the two contributions, so that we can
create nonmaximally entangled states of the form:
\[
| \psi \rangle \propto |H\rangle|H\rangle + \epsilon e^{i
\phi}|V\rangle|V\rangle.
\]
Combined with the single-photon local unitary transformations, any
pure 2-qubit state can be produced.  In this way, we have prepared a
variety of states (Figs.~\ref{TwoQubit}b, \ref{TangleEnt}b ).

\begin{figure}[t]
\epsfxsize=26pc 
\begin{center}
\epsfbox{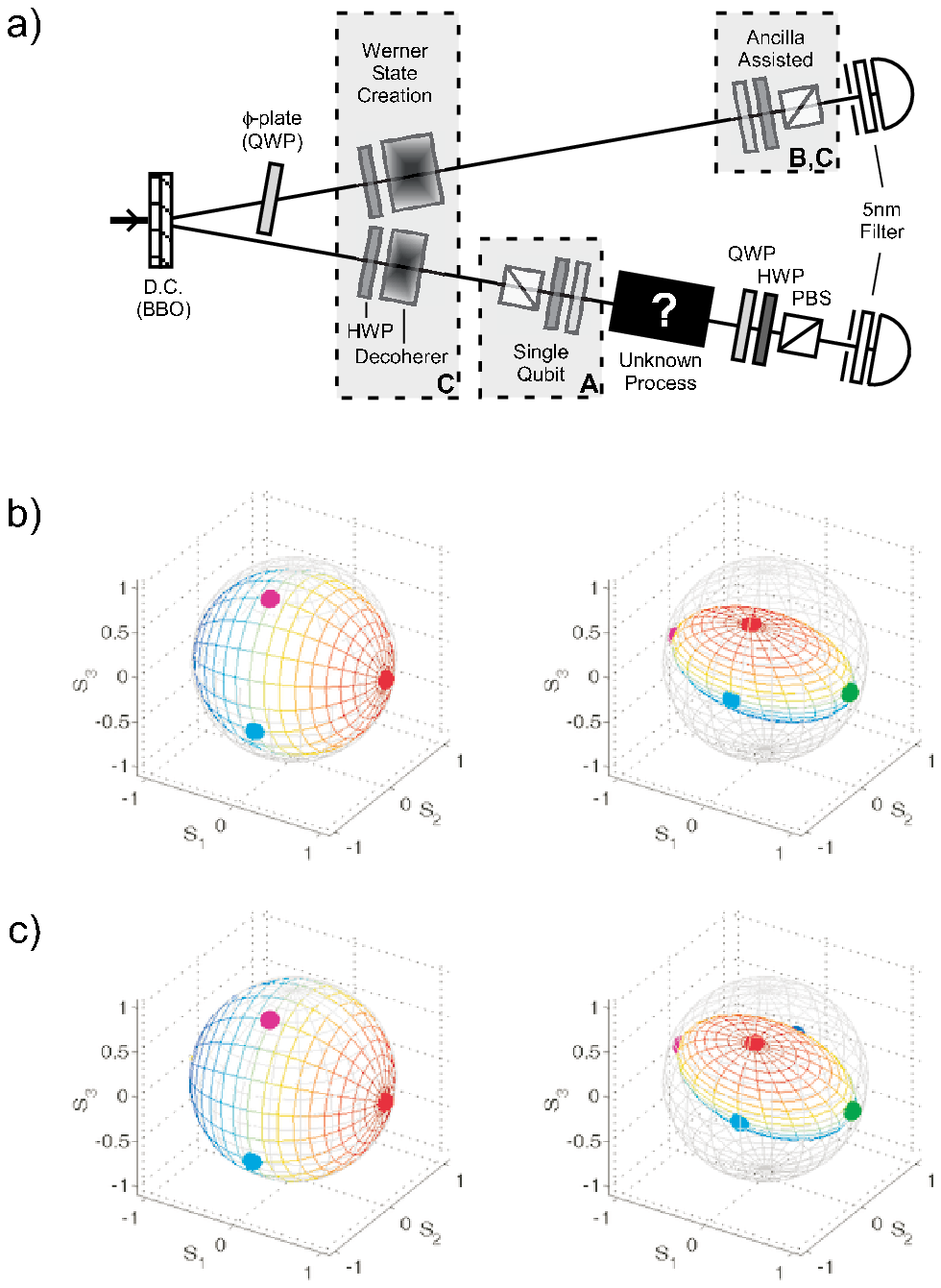} 
\end{center}
\caption{(a) Experimental arrangement for single-qubit and
entanglement-assisted quantum process tomography.  (b) Tomographic
reconstructions for a unitary process (left) and decoherence (right),
illustrated as transformations on the Poincare sphere.  The gray mesh
spheres represent all possible initial single-qubit states.  The dots
show the initial states $|H\rangle$, $|V\rangle$,
$|45^{\circ}\rangle$, $|-45^{\circ}\rangle$, $|L\rangle$, and
$|R\rangle$ after the transformations.  (c) Entanglement-assisted
tomographic reconstructions for the same processes.}
\label{ProcessTomo}
\end{figure}

The density matrices are tomographically
determined by measuring the polarization correlations in 16 bases, and
performing a maximum-likelihood analysis to find the legitimate
density matrix most consistent with the experimental results\cite
{James}. In order to improve the speed and accuracy of our tomographic
measurements, we have implemented a fully automated system.  In
addition to reducing the total time for a measurement, and
significantly decreasing the uncertainty in the measurement settings,
this automated system will also enable the implementation of an adaptive
tomography routine -- by making an initial fast estimate of the state,
one could spend most of the data collection time making an optimized set
of measurements.  With this sort of optimal quantum tomography, we
hope to reach the ultimate limit in quantum state characterization.

Our automated system has enabled the creation of a large
number of states with widely varying degrees of purity and
entanglement.  A convenient way to display these states is the
``Tangle-entropy'' plane, shown in Fig.~\ref{TangleEnt}.  Because it
is impossible to have a state that is both completely mixed and
completely entangled, there is an implied boundary between states that
are physically possible and those that are not: this boundary is
formed by the ``maximally entangled mixed states'' (MEMS), which
possess the largest degree of entanglement possible for their
entropies\cite{T.C.}.

Finally, using the modification of our system shown in
Fig.~\ref{ProcessTomo}a, we can realize several methods of quantum
process tomography\cite{NielsenChuang}, whose goal is to completely 
characterize some
unknown process affecting a qubit. This process may be any
combination of unitary transformations, state-dependent losses, and
decoherence.  One method is to send a variety of input states through
the process, and tomographically determine the output states.  Another
technique, known as ``entangelement-assisted'' or ``ancilla-assisted''
process tomography\cite{Altepeter}, exploits the two-photon
correlations available at the source, and requires only a single,
fixed input state to perform an entire process tomography.

In the future, we will continue to expand our abilities to create an
ever-widening range of quantum states and processes, and to push the
level of precision with which they are created and characterized with
adaptive tomography.  Ultimately, this promising set of tools should
be useful for implementing and testing various protocols in quantum
information processing.

This work was supported in part by the DCI Postdoctoral Research Fellowship
Program and by ARDA, and in part under NSF Grant \#EIA-0121568.


\begin{thebibliography}{99}

\bibitem{Mandel}
C. K. Hong and L. Mandel, \Journal{\PRL}{56}{58}{1986}.

\bibitem{KwiatSource}
P. G. Kwiat {\it et al.}, {\it Phys. Rev. A} {\bf 60}, R773 (1999).

\bibitem{James}
D. F. V. James {\it et al.}, {\it Phys. Rev. A} {\bf 64}, 052312 (2001); A. G.
White {\it et al.}, \Journal{\PRL}{83}{3103}{1999}.

\bibitem{T.C.}
T. C. Wei {\it et al.}, in preparation; W. J. Munro {\it et al.}, {\it
Phys.  Rev.  A} {\bf 64}, 030302(R) (2001).

\bibitem{NielsenChuang}
I. L. Chuang and M. A. Nielsen, {\it J. Mod. Opt.} {\bf 44}, 2455
(1997); J. F. Poyatos {\it et al.}, \Journal{\PRL}{78}{390}{1997}.

\bibitem{Altepeter}
J. B. Altepeter {\it et al.}, in preparation.


\end{thebibliography}
\end{document}